\documentclass[a4paper,fleqn]{cas-dc}

\usepackage[numbers]{natbib}
\usepackage{graphicx}
\usepackage{caption}

\def\tsc#1{\csdef{#1}{\textsc{\lowercase{#1}}\xspace}}
\tsc{WGM}
\tsc{QE}
\tsc{EP}
\tsc{PMS}
\tsc{BEC}
\tsc{DE}

\AtBeginDocument{\setlength{\FullWidth}{\textwidth}}

\begin{document}
\let\WriteBookmarks\relax
\def\floatpagepagefraction{1}
\def\textpagefraction{.001}
\shorttitle{}
\shortauthors{}

\title [mode = title]{Probing Topological Superconductivity of oxide nanojunctions using fractional Shapiro steps}

\author[1,2]{Claudio Guarcello}[orcid=0000-0002-3683-2509]
\ead{cguarcello@unisa.it}

\author[1,2]{Alfonso Maiellaro}[orcid=0000-0001-8539-9564]
\ead{amaiellaro@unisa.it}

\author[3,4]{Jacopo Settino}[orcid=0000-0002-7425-4594]
\ead{jacopo.settino@unical.it}

\author[1]{Irene Gaiardoni}[]
\ead{igaiardoni@unisa.it}

\author[1,5]{Mattia Trama}[orcid=0000-0001-5032-7974]
\ead{mtrama@unisa.it }

\author[1,2]{Francesco Romeo}[orcid=0000-0001-6322-7374]
\ead{fromeo@unisa.it}

\author[1,2,6]{Roberta Citro}[orcid=0000-0002-3896-4759]
\cormark[1]
\ead{rocitro@unisa.it}

\address[1]{Dipartimento di Fisica ``E.R. Caianiello'', Universit\`a di Salerno, Via Giovanni Paolo II, 132, I-84084 Fisciano (SA), Italy}

\address[2]{INFN, Sezione di Napoli, Gruppo collegato di Salerno, Italy}

\address[3]{Dipartimento di Fisica, Università della Calabria, Via P. Bucci Arcavacata di Rende (CS), Italy}

\address[4]{INFN, Gruppo collegato di Cosenza, Italy}

\address[5]{Institute for Theoretical Solid State Physics, IFW Dresden, Helmholtzstr. 20, 01069 Dresden, Germany}

\address[6]{CNR-SPIN, Via Giovanni Paolo II, 132, Salerno, I-84084, Italy}

\cortext[cor1]{Corresponding author}

\begin{abstract}
We theoretically discuss the emergence of fractional Shapiro steps in a Josephson junction created by confining a two--dimensional electron gas at an oxide interface. This phenomenon is induced by an alternating current of proper amplitude and frequency and can be tuned by a magnetic field applied perpendicular to the Rashba spin--orbit axis. The presence of fractional Shapiro steps can be associated with the creation of Majorana bound states at the boundaries of the superconducting leads. Our findings represent a route for the identification of topological superconductivity in non--centrosymmetric materials and confined systems in the presence of spin--orbit interaction, offering also new insights into recently explored frameworks.
\end{abstract}


\maketitle

\section{Introduction}
\label{Sec01}

The AC Josephson effect has proven to be an efficient and sensitive probe for investigating exotic superconductors, offering valuable insights into their properties and aiding in understanding the underlying physics. In particular, in order to explore the anharmonic content of the current--phase relation (CPR) in unconventional systems, recent proposals have pursued tracking of phase--locked behavior in a Josephson junction (JJ) under high--frequency AC drives \cite{Sellier2004,Bae2009,Rokhinson2012,Lee2015,Sau2017,Snyder2018,LeCalvez2019,Panghotra2020,Ueda2020,Kalantre2020,Yao2021,Wang2022,Huang2023,Iorio2023}. In the case of a non--sinusoidal CPR, the AC response exhibits not only the typical constant--voltage plateaus, known as ``integer'' \emph{Shapiro steps} \cite{Shapiro1963}, in the voltage \emph{versus} current characteristic (VIC), but also other steps at specific ``fractional'' voltages, $V_{n/q}=(n/q)\Phi_0 \nu_{\text{ac}}$, with $q\neq n$, $\Phi_0=h/(2e)$ the flux quantum ($e$ is the elementary charge and $h$ is the Planck's constant), and $\nu_{\text{ac}}$ the frequency of the driving signal. Moreover, the $4\pi$--periodicity of Majorana bound states (MBS) spectrum in topological JJs has been speculated to leave traces on the Shapiro response \cite{Wiedenmann2016,Bocquillon2017,Bocquillon2018}. Despite the experimental feasibility of the various approaches developed so far, giving a reliable interpretation of a specific experiment can be complicated due to the nonlinear character of the Josephson response.

Here, we show that fractional AC Josephson effect can also emerge in an oxide--based JJ made by constraining a two--dimensional electron gases (2DEGs), which can form, e.g., at the interface between transition--metal oxides like LaAlO$_3$ and SrTiO$_3$ (LAO/STO), to get a quasi--1D system. Recently, taking a clue from some experimental evidences \cite{Stornaiuolo2017,Tiira2017,Singh2022}, a non--trivial response of the critical current to the magnetic field in this kind of systems has also been theoretically reported \cite{Maiellaro2023}; the latter work discusses also the link to the appearance of MBSs with an ``orbital--flavoured'' internal structure. Here, we demonstrate that the same system can exhibit an unconventional voltage response when driven by an AC signal. In particular, we are referring to the emergence of sizable fractional Shapiro steps in VICs of the junction: those magnetic fields responsible for the increase in critical current, as theoretically argued in Ref. \cite{Maiellaro2023}, also induce the unconventional step structure in the VIC.

The paper is organized as follows: in Sec. \ref{Sec02}, we introduce the microscopic model to obtain the CPRs at different applied magnetic fields; in Sec. \ref{Sec03}, we numerically study the response of our device to an AC drive, i.e., we incorporate the previously obtained CPRs into the so--called resistively shunted junction model, which is commonly used to characterize the time--averaged voltage measured in a current--biased JJ; in Sec. \ref{Sec04}, conclusions are drawn.

\begin{figure*}[t]
\centering
\includegraphics[width=2\columnwidth]{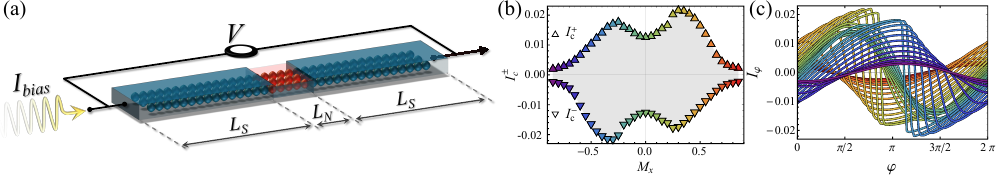}
\caption{(a) Cartoon of the device, including the oscillating current biasing the junction. (b) 
Positive (upward-pointing triangles) and negative (downward-pointing triangles) critical currents as a function of $M_x$. (c) Collection of CPRs obtained by varying $M_x$ from -0.85 (purple curve) to 0.85 (red curve) in increments of 0.05, used to extract the critical currents shown in (b).  The colouring matches that of data in (b).}
\label{Fig00}
\end{figure*}

\section{The microscopic model}
\label{Sec02}

Here, we consider the microscopic theoretical framework treatment presented in Ref. \cite{Maiellaro2023}, i.e., a short superconductor--normal--superconductor (SNS) junction formed by constraining the 2DEG at the LAO/STO (001) interface, see Fig,~\ref{Fig00}(a). The nanochannel Hamiltonian is given by
\begin{equation}\label{Hamiltonian}
 H = H^0 + H^P + H^{SO} + H^{Z} + H^{M},
\end{equation}
where
\begin{eqnarray}
 H^0 \!\!\!\!\!\!&=&\!\!\!\!\!\! \sum_{j} \Psi^{\dagger}_j (h^{0}_{on} \otimes \sigma_0) \Psi_j + \Psi^{\dagger}_j (h^{0}_{hop} \otimes \sigma_0) \Psi_{j+1} + \text{h.c.},\quad\;\, \\\label{HP}
 H^P \!\!\!\!\!\!&=&\!\!\!\!\!\! \sum_{j,\alpha} \Delta(j,q)\ c^{\dagger}_{j,\alpha,\uparrow} c^{\dagger}_{j,\alpha,\downarrow} + \text{h.c.},\\
 H^{SO} \!\!\!\!\!\!&=&\!\!\!\!\!\! \Delta_{SO} \sum_{j} \Psi^{\dagger}_j \left(l_x \otimes \sigma_x + l_y \otimes \sigma_y + l_z \otimes \sigma_z \right) \Psi_j,\\
 H^Z \!\!\!\!\!\!&=&\!\!\!\!\!\! -i \frac{\gamma}{2} \sum_{j} \Psi^{\dagger}_j \left(l_y \otimes \sigma_0 \right) \Psi_{j+1} + \text{h.c.},\\
 H^M \!\!\!\!\!\!&=&\!\!\!\!\!\! M_x \sum_{j} \Psi^{\dagger}_j \left(l_x \otimes \sigma_0 + l_0 \otimes \sigma_x \right) \Psi_j
\end{eqnarray}
using $t_{2g}$ orbitals ($d_{xy}$, $d_{yz}$, and $d_{zx}$) basis. Here, $$\Psi_j=(c_{yz,\uparrow,j},c_{yz,\downarrow,j},c_{zx,\uparrow,j},c_{zx,\downarrow,j},c_{xy,\uparrow,j},c_{xy,\downarrow,j})^T$$ represents the electron annihilation operators for spin, orbital, and position. The terms $H^0$, $H^P$, $H^{SO}$, $H^Z$, $H^M$ correspond to kinetic energy, mean--field pairing, spin--orbit coupling (SOC), inversion--symmetry breaking, and Zeeman interaction, respectively \cite{PhysRevB.100.094526}. The nanochannel is oriented along the $x$--direction \cite{PhysRevB.103.235120}, with the topological phase stabilized by a magnetic field perpendicular to the orbital Rashba--like field \cite{PhysRevB.102.224508,PhysRevB.100.094526}.
$\sigma_i$ are Pauli matrices, while $\sigma_0$ is the identity matrix. $l_x$, $l_y$, $l_z$ are projections of the $L=2$ angular momentum operator onto the $t_{2_g}$ subspace. Analytic expressions of these matrices along with hopping Hamiltonians $h^0_{on}$ and $h^0_{hop}$ are provided in Ref. \cite{Maiellaro2023}. Based on \emph{ab initio} estimates and spectroscopic studies \cite{Pai_2018,PhysRevB.87.161102,Joshua12}, consistent with Ref. \cite{PhysRevB.100.094526}, we assume: $t_1=300$, $t_2=20$, $\Delta_{SO}=10$, $\Delta_t=-50$, and $\gamma=40$ (in meV). $t_1$ and $t_2$ are $x$--directed intraband hopping couplings for the $yz$ and $zx/xy$ bands in $H^0$. $\Delta_t$ denotes the crystal field potential due to symmetry lowering from cubic to tetragonal, and it represents a physical regime with a hierarchy of electronic energy scales such that $|\Delta_t|>\gamma>\Delta_{SO}$.\\
\begin{figure*}[t!!]
\centering
\includegraphics[width=2\columnwidth]{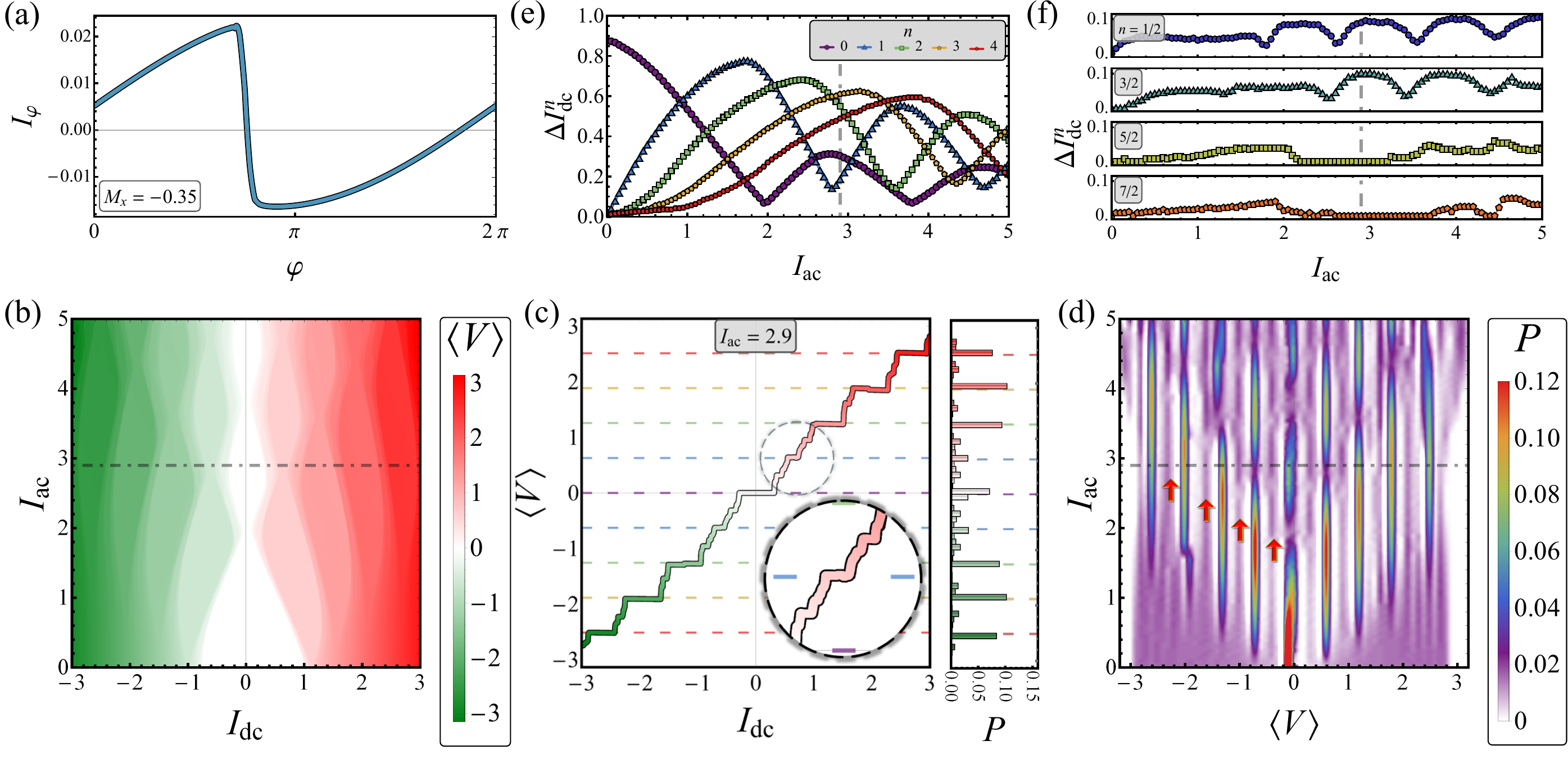}
\caption{Effect of a non--sinusoidal CPR, obtained for $M_x=-0.35$, on the AC Josephson effect, for $I_{\text{dc}}\in[-3,3]$, $I_{\text{ac}}\in[0,5]$, and $\omega_{\text{ac}}=0.1\times2\pi$. (a) CPR. (b) Average voltage drop, $ \left< V\right>$, on the $\left ( I_{\text{dc}},I_{\text{ac}} \right )$--parameter space, and (c) selected $ \left< V\right>$ \emph{vs} $I_{\text{dc}}$ profile at $I_{\text{ac}}=2.9$ (the closeup zooms on the half--integer steps). In the vertical right side of this plot, a histogram, $P\left ( \left< V\right> \right )$, of binned voltage data helps to easily visualize the main step arrangement (the bar height counts the fraction of values lying in each bin). The dashed lines are guides to the eye for the ISSs. (d) Voltage histogram map, $P\left ( \left< V\right>,I_{\text{ac}} \right )$; red arrows mark the peaks associated to the half--integer step. Current step widths, $\Delta I_{\text{dc}}^n$: (e) Integer, i.e., $n=\{0,1,2,3, \text{ and } 4\}$, and (f) half--integer, i.e., $n=\{1/2, 3/2, 5/2, \text{ and } 7/2\}$, as a function of $I_{\text{ac}}$. In panels (b,d,e,f), a gray dot-dashed line marks the value $I_{\text{ac}}=2.9$ chosen for the VICs in (c). Histogram bins have a width of $\delta V=0.1$.} 
\label{Fig01}
\end{figure*}
In Eq.~\eqref{HP}, $\alpha$ is an orbital index, and $\Delta (j,q)=\Delta e^{iqj} e^{i\varphi_j}$ is a space--dependent gap, considering a phase gradient ($\varphi_j=\chi_j^L \varphi_L+\chi_j^R \varphi_R$, with $\chi_j^\alpha=1$ only when $j$ belongs to the $\alpha=L, R$ electrode) induced by the bias current and a finite momentum effect of the Cooper pair ($q$). 
It has been demonstrated that superconductors with both broken inversion and time--reversal symmetries, and in presence of SOS, Cooper's pairs may acquire a finite momentum \cite{doi:10.1073/pnas.2119548119}. Therefore, a spatially modulated superconducting gap is included, with $q$ continuously determined by the magnetic field ($q=\eta M_x$). The pairing amplitude is determined as $\Delta=0.05$, while the phase gradient is $\varphi=\varphi_L-\varphi_R$, with $\varphi_{L,R}$ being the phase of the order parameter of the left and right lead.\\
The Josephson current of a short SNS junction with translational invariant leads is efficiently calculated using the subgap Bogoliubov--de Gennes (BdG) spectrum of a system with truncated superconducting leads: 
\begin{equation}
I(\varphi)=-\frac{e}{\hbar} \sum_{n} \frac{dE_n}{d\varphi},
\end{equation}
where $E_n$ are the subgap energies of the BdG spectrum. The finite--lead approach is valid in the short--junction limit, i.e., $L_N \ll \xi$, with $\xi$ the BCS coherence length. Thus, the total system size is set as $L=2L_S+L_N$, with $L_S \gg \xi$ and $L_N \ll L_S$ the sizes of the superconducting lead and the normal region, respectively. The maximum (absolute value of minimum) of $I(\varphi)$ yields the critical current, $I_c$, in the positive (negative) direction.

The tight--binding Hamiltonian is numerically treated using KWANT \cite{Groth_2014} and solved with NUMPY routines \cite{citeulike:9919912}. We explore different electronic regimes varying the Zeeman energy $M_x$ and the phase difference $\varphi$. 
In this paper, we set $L_S$ to 1000 and $L_N$ to 10, both in units of the lattice constant, and we look only at the orbital filling regime corresponding to the lowest doublet in the energy spectrum with the $d_{xy}$ orbital character.

\section{Results}
\label{Sec03}

Our baseline is the recent experimental evidence of unconventional magnetic--field dependence of the Josephson current discussed in Ref. \cite{Singh2022} and the corresponding microscopic treatment developed in Ref. \cite{Maiellaro2023}. The latter disclosed a connection between topological superconductivity (TSC) and the critical current pattern of a short SNS junction based on non--centrosymmetric superconductors.
Specifically, the hypothesis of a topological phase transition put forward in Ref. \cite{Maiellaro2023} is based on two key observations regarding the critical current of the junction: \emph{i}) the enhancement with increasing applied magnetic field perpendicular to the SOC and \emph{ii}) the distinct symmetry observed when reversing both the magnetic field and the bias current. These features are also clearly highlighted in the magnetic patterns of the critical current shown in Fig.~\ref{Fig00}(b). These characteristics are intertwined with the formation of MBSs at the edges of superconducting conductors and to multiband effects, which relevance can be fine--tuned by appropriate gating of the system \cite{Maiellaro2023}. Thus, a non--trivial topology can manifest in peculiarities of critical current patterns as the magnetic field varies, in line with the experimental findings in Refs. \cite{Tiira2017,Stornaiuolo2017,Singh2022}.

Instead, in this paper we adhere to a different, and at the same time complementary line of thinking, i.e., we demonstrate that even the system's AC response can exhibit non--trivial peculiarities, still assuming working conditions such as those imposed in Ref. \cite{Maiellaro2023}. In particular, we survey the emergence of sizable half--integer Shapiro steps (HISSs) at those applied magnetic fields that enhance the critical current.

In general, the appearance of HISSs is a quite intriguing phenomenon: in the presence of a microwave excitation, the VICs exhibit periodic modulations, known as Shapiro steps \cite{Barone82}, which occur at voltages that are integer multiples of the product of the microwave frequency and $\Phi_0\simeq 2.07\times 10^{-3}\;\text{mV/GHz}$. In unconventional superconductors, which can exhibit non--trivial order--parameter symmetries, HISSs may also appear. These steps can provide valuable information about the nature of the underlying superconducting state.

We first focus on two specific cases, i.e., when the magnetic field is such that: \emph{i}) a topological phase emerges and the critical current was demonstrated to approach a maximum (i.e., we fix $M_x=-0.35$) and \emph{ii}) there are no MBSs and the magnetic field is far from the values that maximize the critical current (i.e., we fix $M_x=0.85$) \cite{Maiellaro2023}.

Through the CPRs resulting from our microscopic analysis (see Fig. \ref{Fig00} (c) for a collection of CPRs obtained for $M_x\in[-0.85,0.85]$), we determine the full AC response of the system by inserting them into a resistively and capacitively shunted junction (RCSJ) model \cite{Barone82}. It describes a current--biased JJ as a parallel arrangement of various current contributions: the external bias current, the Josephson term accounting for Cooper-pair dissipationless flow (i.e., the CPR), a resistive contribution from quasiparticle tunneling, the displacement current, and the thermal current. The use of RCSJ models is a very common practice when studying the AC response of unconventional current--driven Josephson systems.
We will exclusively look at a noise--free overdamped JJ, in which the so--called Stewart--McCumber parameter, $\beta_c = 2eI_c R^2_N C/\hbar$ (here, $R_N$ and $C$ are the normal-state resistance and the capacitance of the junction, respectively), is very small \cite{Stewart1968,McCumber1968,Barone82}, i.e., we neglect the inertia term \footnote{For example, in InAs--nanowire--based JJs for Shapiro step analysis a Stewart--McCumber parameter $\beta_c<0.01$, i.e., an highly overdamped junction, was estimates, thus supporting the use of a resistively shunted junction (RSJ) model \cite{Ueda2020}. In this way, we are also ruling out the effects of the junction capacitance as a possible source of the discussed phenomenology \cite{Hasselberg1975,Valizadeh2008}.  In any case, a critical limit for the Stewart--McCumber parameter exists, i.e., $\beta_c  < 1/4$, below which the inertial contributions to phase dynamics can be safely ignored~\cite{Qian1988,Levi1988}. Fulfilling this condition allows for the omission of the term $C\Phi_0\ddot{\varphi}/(2\pi)$ in the model, thereby also eliminating chaotic behaviors and resulting in a highly damped, more stable system \cite{Strogatz1994,Hilborn2000,Kalantre2020}.}. This constraint is well--suited for externally shunted junctions of any nature and offers a qualitative understanding of (unshunted) weak links \cite{Likharev1988}. 
Thus, by normalizing time to the inverse of plasma frequency, $\omega_p = \sqrt{2\pi I_c/(\Phi_0 C)}$ \cite{Barone82}, the RSJ model, in dimensionless units, can be easily expressed as $I_\varphi+d\varphi/dt=I_{bias}(t)$, where we include both DC and AC bias current components, $I_{bias}(t)=I_{\text{dc}}+I_{\text{ac}}\sin(\omega_{\text{ac}} t)$, and we look te the average voltage drop $ \left< V\right> $ across the junction. For each set of $M_x$, $I_{\text{dc}}$, and $I_{\text{ac}}$, at a given $\omega_{\text{ac}}$, the RSJ model is integrated via a fourth--order Runge--Kutta method over $10^4$ time steps.

\begin{figure*}[t]
\centering
\includegraphics[width=2\columnwidth]{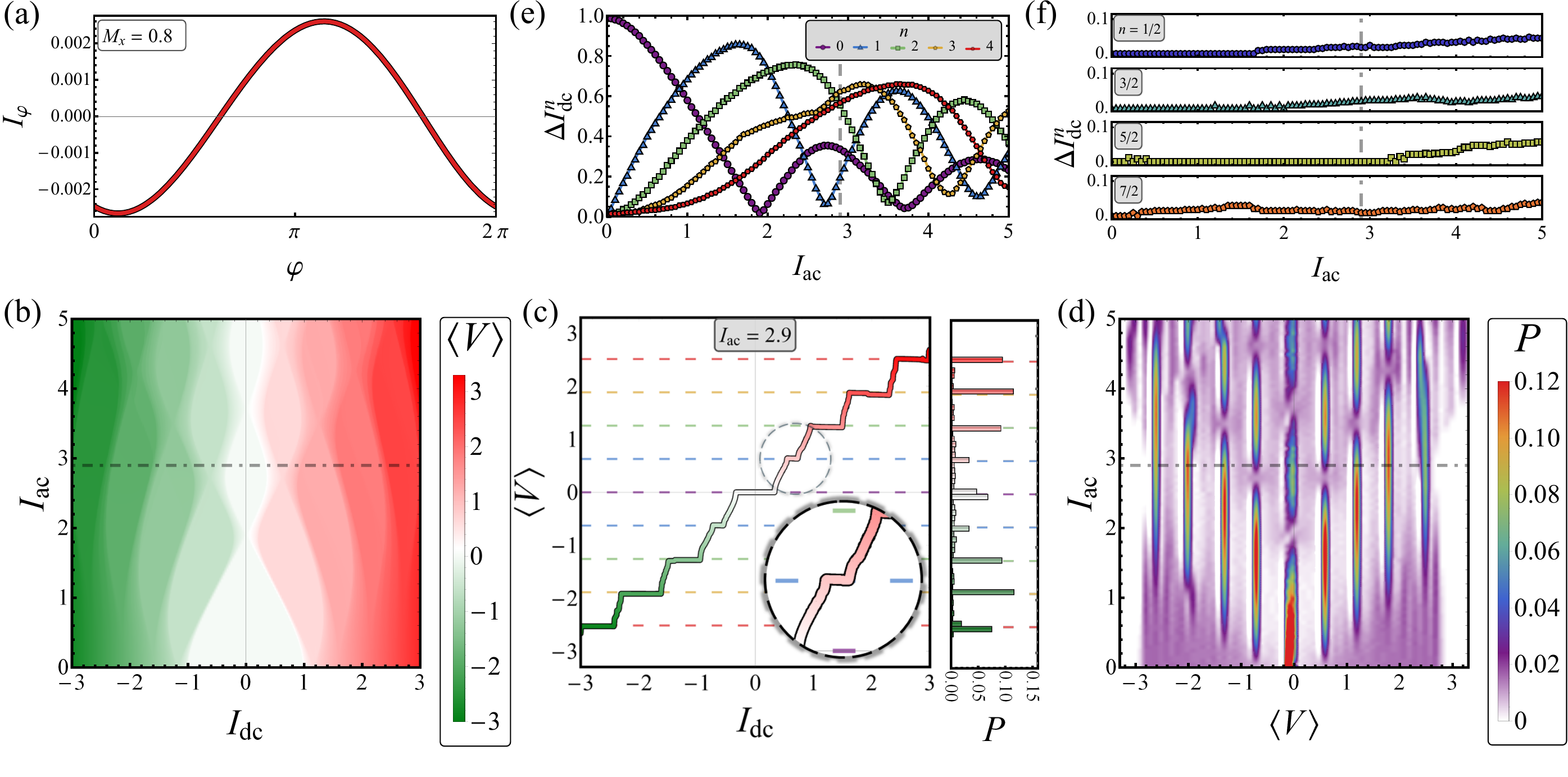}
\caption{Effect of a quasi-sinusoidal CPR, obtained for $M_x=0.8$, on the AC Josephson effect, for $I_{\text{dc}}\in[-3,3]$, $I_{\text{ac}}\in[0,5]$, and $\omega_{\text{ac}}=0.1\times2\pi$. (a) CPR. (b) Average voltage drop, $ \left< V\right>$, on the $\left ( I_{\text{dc}},I_{\text{ac}} \right )$--parameter space, and (c) selected $ \left< V\right>$ \emph{vs} $I_{\text{dc}}$ profile at $I_{\text{ac}}=2.9$ (the closeup zooms on the half--integer steps). In the vertical right side of this plot, a histogram, $P\left ( \left< V\right> \right )$, of binned voltage data helps to easily visualize the main step arrangement (the bar height counts the fraction of values lying in each bin). The dashed lines are guides to the eye for the ISSs. (d) Voltage histogram map, $P\left ( \left< V\right>,I_{\text{ac}} \right )$. Current step widths, $\Delta I_{\text{dc}}^n$: (e) integer, i.e., $n=\{0,1,2,3, \text{ and } 4\}$, and (f) half--integer, i.e., $n=\{1/2, 3/2, 5/2, \text{ and } 7/2\}$, as a function of $I_{\text{ac}}$. In panels (b,d,e,f), a gray dot-dashed line marks the value $I_{\text{ac}}=2.9$ chosen for the VICs in (c). Histogram bins have a width of $\delta V=0.1$.}
\label{Fig02}
\end{figure*}

As already mentioned, the VICs of a JJ can exhibit Shapiro steps, which result from the interaction between external microwave radiation and the intrinsic dynamics of the superconducting phase difference across the junction.
To understand why this phenomenon occurs, according to the RSJ model, we can look at the JJ as a ``phase particle'' in a tilted washboard potential, $U$. 
Under an external microwave drive with frequency $\nu_{\text{ac}}$, the motion of this phase particle synchronizes with the drive. In each excitation period $T_{\text{ac}}$, the phase across the junction changes by $2\pi n$, where $n$ is an integer.
The velocity, $\dot{\varphi}$, of this synchronized motion is given by $2\pi n \nu_{\text{ac}}$. Remarkably, this synchronization is stable within certain intervals of the applied dc--current, $I_{\text{dc}}$, i.e., small changes in $I_{\text{dc}}$ do not disrupt the synchronization, allowing the system to maintain a constant velocity.
This stability results in the appearance of plateau, i.e., the Shapiro steps, in the VICs. Each step corresponds to a constant voltage given by $V_n=\frac{\Phi_0}{2\pi} \dot{\varphi}_n = n\Phi_0 \nu_{\text{ac}}$, with $n$ integer, so that Shapiro steps are equidistant, with a separation $\Delta V = V_{n+1} - V_n = \Phi_0 \nu_{\text{ac}}$. In other words, only the frequency $\nu_{\text{ac}}$ determines the spacing between steps, e.g., at $\nu_{\text{ac}}\approx483.6$ GHz the separation is $\sim1\;\text{mV}$. In the following, we impose $\omega_{\text{ac}}=0.1\times2\pi$, knowing that the overall effect of the drive frequency is to increase the range of DC currents between two successive integer steps as $\omega_{\text{ac}}$ increases \cite{Raes2020}. 

In Fig. \ref{Fig01} we collect the AC response of the system by setting $M_x=-0.35$. In this case, the obtained CPR deviates significantly from a sinusoidal profile, see Fig. \ref{Fig01}(a); it also features an anomalous Josephson contribution \cite{Buzdin2008,Strambini2020,Guarcello2021}, i.e., a non--zero current at zero phase, and a marked asymmetry, i.e., quite different maximum and minimum currents -- the so--called \emph{Josephson diode effect} \cite{Davydova2022,Nadeem2023,Maiellaro2024}. In Fig. \ref{Fig01}(b) we show how the average voltage drop, $ \left< V\right>$, changes on the $\left ( I_{\text{dc}},I_{\text{ac}} \right )$--parameter space; this is a \emph{Shapiro map}, commonly used to chart the widths of the observed plateaus, illustrating their dependence on both the AC drive amplitude and DC current bias (often, in the place of the voltage, the differential resistance, $d V/dI_{\text{dc}}$, is plotted). The formation of plateaus due to integer Shapiro steps (ISSs) is quite evident, just like the transitions between steps, which follow a typical pattern. A selected VIC is presented in Fig. \ref{Fig01}(c), i.e., the $ \left< V\right>$ \emph{vs} $I_{\text{dc}}$ profile for $I_{\text{ac}}=2.9$. The dashed lines are guides to the eye for the ISSs. This plot reveals the formation of additional small steps in the midpoints between integer steps, as also better highlighted in the closeup. Step formation can be appreciated not only directly in the VIC, but also by binning the measurement data according to voltage, i.e., through a histogram such the one placed vertically to the right of Fig. \ref{Fig01}(c). Here, the bar height counts the fraction of values lying in each bin, so that each peak marks to an ISS. Further information can then be gathered by taking into account all $P\left ( \left< V\right> \right )$ histograms as $I_{\text{ac}}$ varies, in order to construct a \emph{voltage histogram map}, $P\left ( \left< V\right>,I_{\text{ac}} \right )$, as in Fig. \ref{Fig01}(d). This density plot shows the ISSs very clearly, but we can also resolve ``islands'' of more intense color amidst the ISSs, which attest to the formation of HISSs (also pointed out by red arrows). Furthermore, a careful look reveals that, especially for high $I_{\text{ac}}$ values, other fractional resonances tend to emerge, corresponding to phase locking with different higher--order CPR harmonics. However, these resonances are significantly narrower than their half--integer counterparts, making them potentially more vulnerable to suppression by thermal fluctuations. Indeed, temperature is a quite important parameter to consider, as it has been observed that the height of the Shapiro steps can nonmonotonically depende on it~\cite{Revin2021}.

The width $\Delta I_{\text{dc}}^n$ of the ISSs, $n=\{1,2,3,4, \text{ and } 5\}$, and HISSs, $n=\{1/2, 3/2, 5/2, \text{ and } 7/2\}$, by changing $I_{\text{ac}}$ is reported in Fig. \ref{Fig01}(e)--(f). In both cases, we observe a pronounced Shapiro response. At zero current, there is only the $n=0$ step; then, as $I_{\text{ac}}$ grows, other integer steps ``switch on", until the lobe structure typical of ISSs emerges, with the step widths in current exhibiting an oscillating Bessel--function--like pattern \cite{Russer1972}. Even for HISSs, we can appreciate the formation of lobes in the $\Delta I_{\text{dc}}^n(I_{\text{ac}})$ profile.

\begin{figure}[t!!]
\centering
\includegraphics[width=\columnwidth]{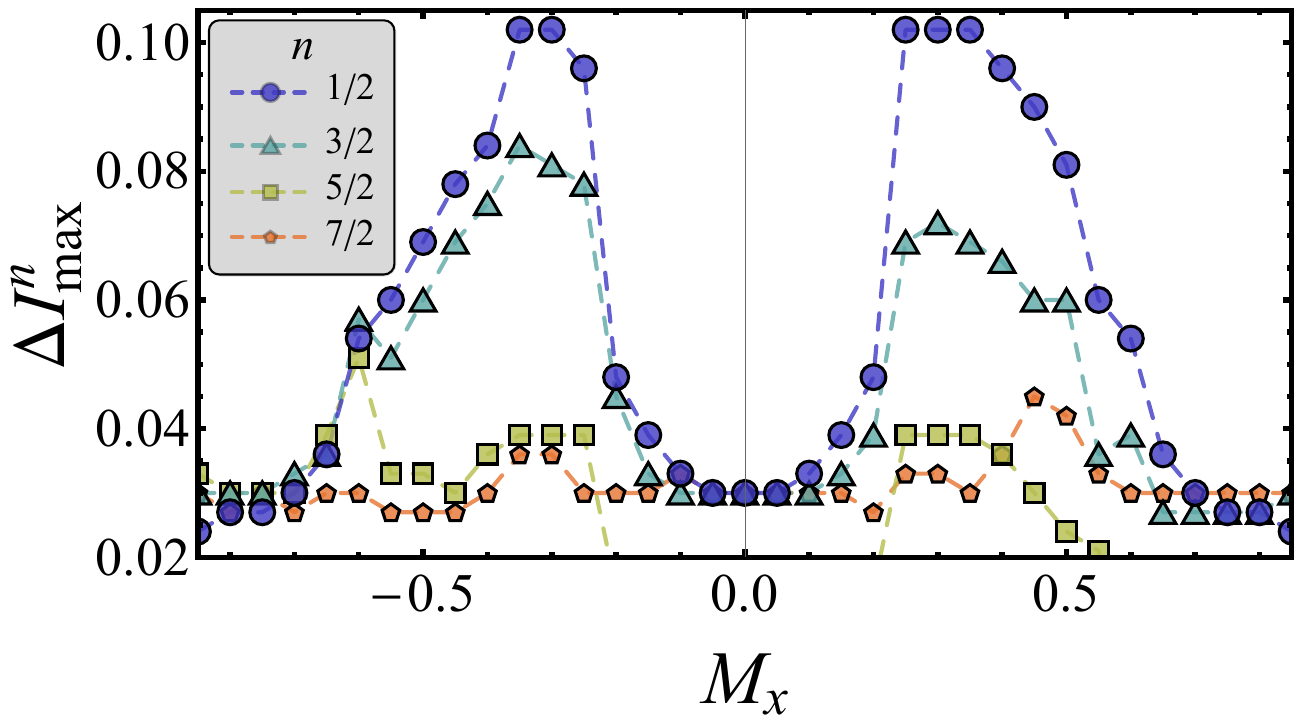}
\caption{Magnetic-field dependence of the HISSs maximum width, $\Delta I_{\text{max}}^n$ with $n=\{1/2, 3/2, 5/2, \text{ and } 7/2\}$.} 
\label{Fig03}
\end{figure}

The situation changes if we consider a magnetic field giving a CPR closer to a sinusoidal profile, see Fig. \ref{Fig02}(a) for $M_x=0.85$. The $ \left< V\right>\left ( I_{\text{dc}},I_{\text{ac}} \right )$ Shapiro map still shows the formation of plateaus, see Fig. \ref{Fig02}(b), but if we look in detail at a VIC, see Fig. \ref{Fig02}(c), it is evident that the HISSs do not form so clearly this time. This is confirmed by the histogram map in Fig. \ref{Fig02}(d). If we graph the step width behavior as $I_{\text{ac}}$ changes, we observe that the ISSs again follow a lobe structure, which is however lost when we look at the HISSs (in the latter case, there is only an increasing trend due to the formation of spurious tones in the VIC as $I_{\text{ac}}$ increases), see Fig. \ref{Fig02}(e-f). We note also that the width of the integer steps at the minimum intralobes is non--zero, and the comparison of Figs. \ref{Fig02}(e) and \ref{Fig03}(e) reveals that it tends to increase as the skewness of the CPR rises; this agrees with the fact that a residual non--sinusoidal correction of the CPR leads to an integer step width that does not completely cancel out changing the AC amplitude and frequency \cite{Raes2020} (this phenomenon is well illustrated in the animations in the Supplementary Material).

To get an overall insight into the effects induced by the external magnetic field on the Shapiro response, we collect in Fig. \ref{Fig03} the maximum step width, $\Delta I_{\text{max}}^n =\max_{I_{\text{ac}}}\left\{\Delta I_{\text{dc}}^n\right\}$, of the HISSs for $M_x\in[-0.85,0.85]$. Especially when looking at the steps with $n=1/2$ and $3/2$, the formation of two peaks standing out clearly against a small, non--zero background is quite evident. These peaks form just at those $M_x$'s at which the enhancement of the critical current is observed, see Fig. \ref{Fig00}(b), namely, the magnetic field values that were pointed out as being responsible for the emergence of topological features \cite{Maiellaro2023}. 
In other words, we assert that those values of $M_x$ for which the system enters a topological phase, not only give a maximum of the critical current, but also should induce measurable half--integer steps in the VIC of a junction excited by an AC drive.

\section{Conclusions}
\label{Sec04}

In this paper, we extended the theoretical findings in Ref. \cite{Maiellaro2023} showing that the emergence of half--integer Shapiro steps can be a valuable tool for recognizing topological signatures in Josephson junctions based on non--centrosymmetric superconductor. 
Numerically calculating the AC response of this system, we showed that half--integer Shapiro steps occur clearly within a specific magnetic field range of values; in particular, the phenomenon takes place at the magnetic fields giving the increase in critical current that was reported in Ref. \cite{Maiellaro2023} in connection with the emergence of Majorana modes at the edges of superconducting leads. We observe that the visibility of the half--integer Shapiro steps scales with the external magnetic field showing a very peculiar two--lobe structure. These findings provide a tool for proving the presence of topological phase \cite{Guarcello2020}, complementary to that discussed in Ref. \cite{Maiellaro2023}, which could be promptly employed in, e.g., LAO/STO--type \cite{Stornaiuolo2017,Singh2022} or nanowire--based systems \cite{Tiira2017,Wu2024}. The appearance of a fractional Shapiro response marks a deviation from topologically--trivial to nontrivial junctions, potentially driven by the external magnetic field.

Finally, we conclude with a word of caution, since the point of the experimental feasibility is indeed delicate and we stress a constraint of our proposal: in fact, the structure of half--integer steps can also be affected by non--adiabatic effects \cite{Dubos2001,Basset2019}, as the current--phase relation may suffer additional distortions, especially at high driving frequencies, e.g., see Ref \cite{Iorio2023}. In particular, 
the frequency window defining the adiabatic regime should be identified paying special attention to the crossover frequency with the non-adiabatic regime,
which is related to the Thouless energy, in the diffusive case, or the $RI_{\text{SW}}$ product, for ballistic systems (with $I_{\text{SW}}$ being the JJ switching current) \cite{Ueda2020}. Furthermore, even within the adiabatic driving regime, deviations in the current--phase relation can emerge due to the interference of multiple supercurrent paths according to the basic mechanism presented in Ref. \cite{Romeo2004}. This is likely to occur when a 2DEG structure is gated at low density, a condition under which the interaction of the electron density with unavoidable impurities may induce percolation and interference of phase--coherent superconducting paths.
We also note that the occurrence of fractional Shapiro steps has also been attributed to causes not directly related to a non-sinusoidal current-phase relation. For example, in YBCO long Josephson junctions, fractional Shapiro steps have been associated with chaotic dynamics~\cite{Revin2018}.
For these reasons, a comparison with a concrete experiment should certainly take into account the working regimes, and possibly consider a revision of the microscopic model \cite{Iorio2023}. Under the mentioned limits, our approach well captures the key aspects of the dynamic Josephson response relevant to our considerations.

\section{Acknowledgments}
\label{Sec05}
This work is financed by Horizon Europe EIC  Pathfinder under the grant IQARO number 101115190.
F.R. acknowledges funding from Ministero dell’Istruzione, dell’Università e della Ricerca (MIUR) for the PRIN project  STIMO (Grant No. PRIN 2022TWZ9NR). M.T. and A.M. acknowledge financial support from "Fondazione Angelo Della Riccia". R.C. acknowledges funding from Ministero dell’Istruzione, dell’Università e della Ricerca (MIUR) for the PRIN project  QUESTIONS (Grant No. PRIN P2022KWFBH). 

\printcredits

\bibliographystyle{cas-model2-names}


\end{document}